\documentclass[]{spie}  %>>> use for US letter paper
\usepackage{amsmath,amsfonts,amssymb}
\usepackage{graphicx}
\usepackage{booktabs}
\usepackage[colorlinks=true, allcolors=blue]{hyperref}

\title{NEXT: The Netherlands EXoplanet Testbed\\ I. Goals and opto-mechanical design}

\author[a,b]{Rico Landman}
\author[b,c]{Sebastiaan Haffert}
\author[b]{Louis Desdoigts}
\author[b]{Matthijs Mars}
\author[b]{Dhwanil Patel}
\author[b]{Daniel Smith}
\author[a,b]{Remko Stuik}
\author[b]{Adam Taras}
\author[b]{Elena Tonucci}
\author[b]{Yinzi Xin}

\affil[a]{NOVA, Netherlands Research School for Astronomy, P.O. Box 9513, 2300 RA Leiden, The Netherlands}
\affil[b]{Leiden Observatory, Leiden University, P.O. Box 9513, 2300 RA Leiden, The Netherlands}
\affil[c]{Steward Observatory, University of Arizona, 933 N. Cherry Ave., Tucson, AZ 85721, USA}

\authorinfo{Send correspondence to R.\ Landman: E-mail: rlandman@strw.leidenuniv.nl}

\begin{document}
\maketitle

\begin{abstract}
We report on the goals, design, and ongoing development of the Netherlands EXoplanet Testbed (NEXT), a high-contrast imaging testbed under construction in Leiden. NEXT is designed to develop and validate technologies and algorithms for extreme adaptive optics (XAO) and coronagraphy at the performance levels required by the next generation of high-contrast imagers on the Extremely Large Telescopes (ELTs) and future space observatories. All powered optics in the common path are custom off-axis parabolas, making the bench fully reflective up to the science cameras, and it operates from the visible to the near-infrared (goal of 500--1800 nm). It combines a woofer-tweeter XAO module, using an ALPAO woofer and a Boston Micromachines kilo-DM as tweeter, with a coronagraphic arm that supports common coronagraph architectures and focal-plane wavefront control for dark-hole digging, and reserves space for a suite of wavefront sensors. The design targets a raw contrast of $10^{-7}$ (goal: $10^{-8}$) at $5\,\lambda/D$ and 800 nm. We present the opto-mechanical design of the testbed, the key trade-offs made to reach these contrast levels, and Fresnel-propagation simulations of its predicted performance.
\end{abstract}


\section{INTRODUCTION}
\label{sec:intro}

The next generation of high-contrast imaging instruments on the Extremely Large Telescopes (ELTs) and from space aim to detect and spectroscopically characterize rocky exoplanets in the habitable zones of nearby stars. Detecting these faint companions in reflected light requires overcoming extreme contrast ratios at small angular separations. To meet this goal, upcoming ELT instruments such as the Planetary Camera and Spectrograph (PCS)\cite{kasper2021} and GMagAO-X\cite{males2024gmagaox}, as well as the future NASA flagship Habitable Worlds Observatory (HWO)\cite{feinberg2024}, demand unprecedented performance from their starlight suppression systems. Achieving these contrast levels requires high-performance extreme adaptive optics (XAO), near-optimal coronagraphs, and focal-plane wavefront control, all of which must be developed and validated before deployment on astronomical instruments.

Laboratory testbeds play a critical role in this development by providing controlled environments in which new hardware, algorithms, and calibration strategies can be tested before on-sky implementation. Existing high-contrast imaging testbeds generally fall into two categories: XAO-focused platforms aimed at developing high-speed wavefront sensing and control, and coronagraphic testbeds optimized for demonstrating deep starlight suppression. Together, they have enabled many of the technologies now being adopted for future ELT instruments and space missions.

For ground-based high-contrast imaging research, several XAO testbeds focus on high-speed, high-order wavefront control together with moderate-contrast coronagraphy. In the US, the Santa Cruz Extreme AO Laboratory (SEAL) provides a flexible platform for testing advanced XAO control and wavefront-sensor (WFS) technologies, and has recently undergone a fully reflective, multi-wavelength rebuild\cite{jensenclem2021, jensenclem2025}. MagAO-X co-functions as both an active, high-order laboratory testbed and a deployed on-sky instrument at the Magellan Clay Telescope\cite{males2018, males2022, males2024}. It serves as a bridge between laboratory demonstrations and astronomical observations, validating high-speed wavefront control, coronagraphy, and science-instrument concepts on-sky. SCExAO fulfills a similar role at the Subaru Telescope\cite{Jovanovic2015PASP..127..890J}. At ESO, the GPU-based High-order adaptive Optics Testbench (GHOST) is a refractive XAO testbed used to develop and benchmark novel wavefront-control methods, including dual-stage XAO and predictive control\cite{engler2024}.

Dedicated coronagraph testbeds instead prioritize deeper contrast by minimizing the number of optical components and maximizing stability. The vacuum-enclosed High Contrast Imaging Testbed (HCIT) and its Decadal Survey Testbeds (DST and DST2) at JPL have demonstrated raw contrasts approaching $10^{-10}$ using pair-wise probing and Electric Field Conjugation\cite{meeker2021, noyes2023}. HiCAT at STScI focuses on wavefront control for complex, segmented, and obstructed apertures\cite{ndiaye2013, ndiaye2015}, while the Space Coronagraph Optical Bench (SCoOB) in Arizona develops coronagraphy and wavefront sensing for future space missions with an emphasis on compactness and stability\cite{ashcraft2022}. In Europe, the THD2 bench at Paris Observatory routinely achieves contrasts of $10^{-8}$ to $10^{-9}$ using multi-deformable-mirror architectures in ambient air\cite{baudoz2018, laginja_spie_2026}, while the SPEED bench in Nice explores high-contrast imaging strategies for segmented apertures, particularly in the context of the ELT\cite{martinez2022, martinez2023}.

To complement these existing international facilities and provide dedicated regional access, we are developing the Netherlands EXoplanet Testbed (NEXT), a flexible platform for XAO and coronagraph technology development. NEXT is optimized to push the boundaries of ground-based XAO by combining high-order wavefront control with advanced coronagraphs at performance levels relevant for PCS. At the same time, it provides a platform for early demonstrations of technologies that are also relevant for future space missions. NEXT has been designed to be compatible with MagAO-X: the coronagraphic arm matches the MagAO-X pupil size and focal ratio, and the control software is based on the same software suite. Components and algorithms can therefore be developed and validated on NEXT before being deployed on-sky.

In this paper we present the goals and design of NEXT. Section \ref{sec:goals} describes the top-level science requirements. Section \ref{sec:design} presents the opto-mechanical design of the testbed and its subsystems. Section \ref{sec:performance} presents the predicted performance from Zemax and Fresnel-propagation simulations, including image quality, throughput, and contrast limits. Section \ref{sec:software} outlines the control software architecture, and Section \ref{sec:outlook} concludes with the current project status and schedule.

\section{GOALS \& REQUIREMENTS}
\label{sec:goals}
The main goal of the testbed is to test novel technologies and algorithms for extreme adaptive optics and coronagraphy at the performance levels required by PCS. Since the primary focus is on maturing technologies for the ELT and PCS, NEXT is architecturally similar to the ground-based XAO testbeds described in Section \ref{sec:intro}. The resulting top-level requirements are summarized in Table \ref{tab:requirements}. The most stringent of these is to reach a raw contrast of $10^{-7}$ (and a goal of $10^{-8}$) at $5\,\lambda/D$ in a one-sided dark zone after focal-plane wavefront control, ideally over the full PCS wavelength range (500--1800 nm) and at a throughput high enough that wavefront control algorithms can be tested quickly and in real time.

\begin{table}[h!]
    \centering
    \def\arraystretch{1.3}
    \begin{tabular}{l c c l}
        \toprule
        \textbf{Top-level requirement} & \textbf{Minimum} & \textbf{Goal} & \textbf{Set by} \\
        \midrule
        Spectral bandwidth & 600--1800 nm & 500--1800 nm & PCS bandwidth \\
        Field of view & $200\,\lambda/D$ & $300\,\lambda/D$ & High-order DM \& WFS \\
        Raw contrast at $5\,\lambda/D$$^*$ & $10^{-7}$ & $10^{-8}$ & Coronagraph \& FPWFS \\
        Photon flux & 0 mag star on ELT & $-5$ mag star on ELT & Dark-hole digging time \\
        \bottomrule
    \end{tabular}
    \vspace{6pt}
    \caption{Top-level requirements for the testbed. $^*$ Raw one-sided contrast after EFC at $5\,\lambda/D$ in a 10\% bandwidth around 800 nm.}
    \label{tab:requirements}
\end{table}
Additionally, the testbed is planned to be used for initial concept demonstrations and component validation for future space-based observatories such as HWO, both in the visible and near-infrared. However, to show contrast levels of $10^{-10}$, final tests are expected to require a thermal-vacuum testbed.

Beyond the contrast requirement, the testbed should be able to accommodate at least the following:
\begin{enumerate}
    \item A turbulence simulator.
    \item A Woofer-tweeter common-path DM architecture for large dynamic range and high actuator count wavefront control.
    \item Fourier-filtering WFS such as the Pyramid wavefront sensor.
    \item A non-common path DM for focal plane wavefront control and dark hole digging.
    \item Common coronagraph architectures (pupil apodizer, PIAA lenses, focal plane mask and Lyot stop).
    \item Sufficient space for WFS using the light rejected by the focal plane mask (FLOWFS) and Lyot stop (LLOWFS).
    \item Sufficient space for science cameras and visitor experiments.
    \item Allows for testing of components and algorithms for deployment on MagAO-X.
\end{enumerate}

\section{OPTO-MECHANICAL DESIGN}
\label{sec:design}
\begin{figure}
    \centering
    \includegraphics[width=\linewidth]{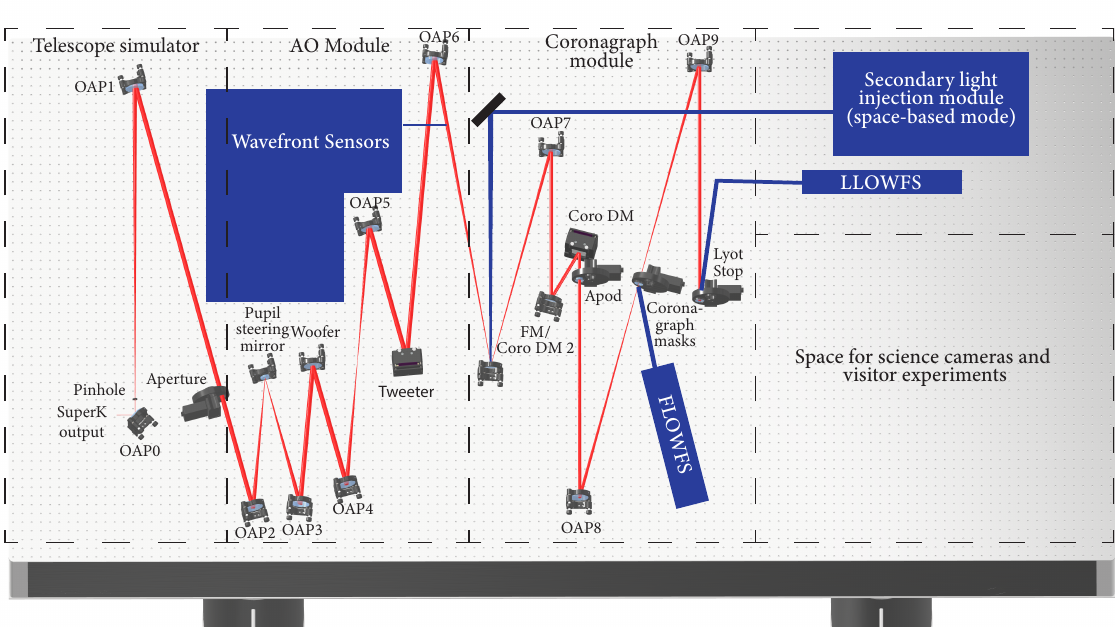}
    \caption{Overview of the optical design of the testbed, which features an AO module, a coronagraph module, and ample space for science cameras and visitor experiments. Abbreviations: CDM: Coronagraphic DM, DM: Deformable Mirror, FM: Fold Mirror, (F/L)LOWFS: focal-plane/Lyot-stop low-order wavefront sensor, OAP: Off-Axis Parabola, PSM: Pupil-Steering Mirror, TS: Turbulence Simulator.}
    \label{fig:overview}
\end{figure}

The testbed layout can be separated into a few distinct modules: a telescope simulator, an XAO module, a coronagraphic module and a scientific backend. The optical design follows the principles established by existing high-contrast imaging benches, balancing the need for enough accessible pupil and focal planes against the size constraints and their impact on throughput and contrast. We discuss the opto-mechanical design and trade-offs for each module in the following subsections. An overview of the full optical layout is shown in Fig. \ref{fig:overview}. The bench is planned to be operable in two modes, which we refer to throughout as the ground-based mode, in which the light traverses the full XAO module before entering the coronagraph, and the space-based mode, in which a secondary telescope simulator is injected directly ahead of the coronagraphic module. The two are selected with a fold mirror at the ground/space switch listed in Table \ref{tab:planes}.

The design started from a first-order, spreadsheet-level layout that was subsequently imported into Zemax and iterated. Three quantities were fixed from the start. The first is the size of the optical table, a Thorlabs T1530 (1.5 $\times$ 3 m). The second is the size of the deformable-mirror clear apertures, which are 13.5 mm and 13.2 mm for the woofer (ALPAO) and tweeter (Boston Micromachines (BMC) kilo-DM) respectively, and 10 mm (Fraunhofer) for the coronagraphic DM, with the option of installing a second kilo-DM as a second coronagraphic DM. The third is the pupil size (9 mm) and focal ratio (F/69) of the coronagraphic arm, chosen for compatibility with MagAO-X\cite{males2018}. To accommodate the DM apertures we adopt a 12.6 mm pupil in the XAO part of the testbed, which maximizes the number of illuminated actuators on the kilo-DM, and demagnify this to 9 mm for the coronagraphic part. Throughout the instrument we use custom off-axis parabolic (OAP) mirrors for all powered optics, in order to obtain diffraction-limited imaging over the full field of view and the entire wavelength range. To minimize manufacturing costs, the custom powered optics are drawn from only three parent OAP prescriptions (350, 621 and 870 mm), and are all mounted in standard 2-inch mounts; the only exception is the 25 mm OAP in the telescope simulator, which is an off-the-shelf 1-inch component. Table \ref{tab:oaps} summarizes the OAP prescriptions and their functions, and Table \ref{tab:planes} lists the key pupil and focal planes of the design.

\begin{table}[h!]
    \centering
    \def\arraystretch{1.2}
    \begin{tabular}{l c c l}
        \toprule
        \textbf{OAP} & \textbf{Parent focal length} & \textbf{Reflection angle} & \textbf{Function} \\
        \midrule
        OAP0 & 25 mm & 90$^\circ$ & Focus the fiber output onto the pinhole \\
        OAP1 & 870 mm & 15$^\circ$ & Collimate the pinhole onto the entrance aperture \\
        OAP2 & 350 mm & 20$^\circ$ & Focus onto the pupil-steering mirror \\
        OAP3 & 350 mm & 20$^\circ$ & Collimate; woofer pupil \\
        OAP4, OAP5 & 350 mm & 20$^\circ$ & Pupil relay onto the tweeter \\
        OAP6 & 870 mm & 15$^\circ$ & Pupil demagnification (with OAP7) \\
        OAP7 & 621 mm & 15$^\circ$ & Form the 9 mm coronagraph pupil \\
        OAP8 & 621 mm & 15$^\circ$ & Focus to the coronagraphic focal plane (F/69) \\
        OAP9 & 621 mm & 15$^\circ$ & Collimate to the Lyot pupil \\
        \bottomrule
    \end{tabular}
    \vspace{6pt}
    \caption{Off-axis parabolas in the testbed. All OAPs are custom components drawn from three parent prescriptions, except OAP0, which is an off-the-shelf 1-inch OAP.}
    \label{tab:oaps}
\end{table}

\begin{table}[h!]
    \centering
    \def\arraystretch{1.2}
    \begin{tabular}{l l l l}
        \toprule
        \textbf{Plane} & \textbf{Type} & \textbf{Element} & \textbf{Size / focal ratio} \\
        \midrule
        Entrance pupil & Pupil & Aperture stop / pupil masks & 12.6 mm \\
        Pupil steering mirror & Focal & Superflat steering mirror & F/28 \\
        Woofer & Pupil & ALPAO DM-97 (13.5 mm) & 12.6 mm \\
        Tweeter & Pupil & BMC kilo-DM (13.2 mm) & 12.6 mm \\
        Ground/space switch & Focal & Superflat fold-mirror & F/69 \\
        Coronagraph pupil & Pupil & CDMs (Fraunhofer, 10 mm) / apodizer / PIAA lenses & 9 mm \\
        Coronagraphic masks & Focal & Focal-plane mask & F/69 \\
        Lyot pupil & Pupil & Inverse PIAA lenses / Lyot stop & 9 mm \\
        \bottomrule
    \end{tabular}
    \vspace{6pt}
    \caption{Key pupil and focal planes of the testbed.}
    \label{tab:planes}
\end{table}

\subsection{Telescope simulator}
The light source is a NKT SuperK COMPACT, a supercontinuum white-light laser covering approximately 450--2400 nm. This source provides sufficient spectral flux density across the full testbed wavelength range. The output of the SuperK fiber is a collimated Gaussian beam with a beam waist of $\sim$1.5 mm at a wavelength of 800 nm. This beam is focused onto a pinhole by a 25 mm off-the-shelf OAP to spatially filter the beam, removing higher-order modes and ensuring that we are simulating a point source. The filtered light is subsequently collimated by an 870 mm parent focal length OAP, and we define a pupil of 12.6 mm diameter at the aperture stop, which acts as the entrance pupil for the remainder of the system. A filter wheel with aperture stops is placed at this pupil, which will at least include aperture masks for the Magellan-Clay telescope and the ELT. A turbulence simulator will be placed as close as possible to the aperture stop. While placing the turbulence simulator outside the pupil will result in some amplitude errors, we have deemed this to be a better solution than adding another pupil relay.

\begin{figure}
    \centering
    \includegraphics[width=\linewidth]{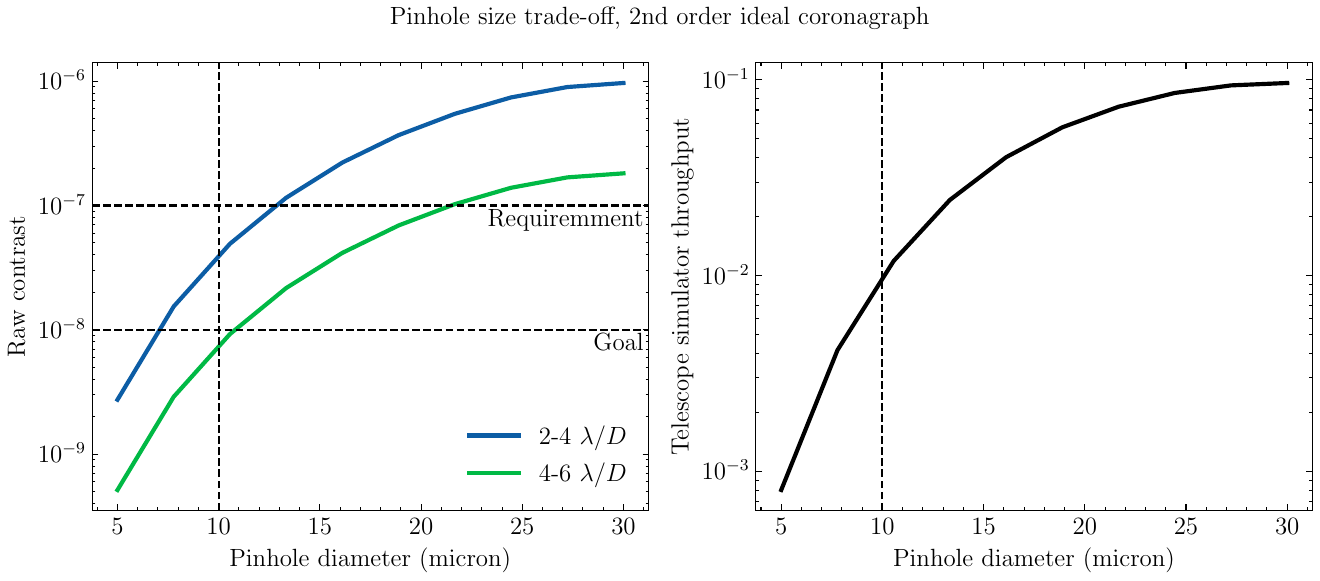}
    \caption{Trade-offs between throughput and contrast as a function of the size of the pinhole at a wavelength of 800 nm. Left: raw contrast as a function of pinhole diameter. Right: throughput of the telescope simulator as a function of pinhole diameter. Simulated with a second-order ideal coronagraph, a 25 mm OAP before the pinhole and an 870 mm collimating OAP.}
    \label{fig:telescope_sim}
\end{figure}

We have performed a trade study of the pinhole size. For a small pinhole, the throughput of the setup is low, which increases dark-hole digging times and results in a larger relative contribution from the thermal background in the near-infrared. On the other hand, too large a pinhole limits the achievable contrast, because the source is then no longer effectively unresolved: the finite angular extent of the pinhole is not fully rejected by the coronagraph and leaks into the dark zone. We simulated the telescope simulator in \texttt{hcipy}\cite{por2018} using an ideal second-order coronagraph. Fig. \ref{fig:telescope_sim} shows the resulting throughput and raw contrast as a function of pinhole diameter. For diameters $\lesssim 13$ $\mu$m, raw contrasts $< 10^{-7}$ can be achieved for separations $> 2\,\lambda/D$, while larger pinholes compromise performance. We have settled on a $\sim$10 $\mu$m pinhole, which offers a good balance between throughput and contrast. Dark-hole digging algorithms may further reduce the impact of the finite pinhole size. Alternatively, a larger pinhole could be swapped in for higher throughput in the near-infrared.

Because the SuperK is a class-IIIb source, the telescope simulator up to the aperture stop will be enclosed in a dedicated laser-safety box; the residual power downstream of the aperture stop is low enough for the beam to be handled safely with visible light filters. A HeNe laser is foreseen for alignment purposes as an alternative to the white-light source.

A secondary telescope simulator is planned to be added later, injected into the main beam path right before the coronagraphic module. It provides a coronagraph-only mode with a shorter beam path that avoids the accumulation of wavefront and amplitude errors from the (deformable) mirrors in the AO module, and should therefore allow for deeper contrasts. This is the space-based mode referred to in Section \ref{sec:fresnel}.

\subsection{Adaptive optics module}
After the telescope simulator, the beam is focused by a 350 mm OAP onto a pupil-steering mirror (PSM), which is used to align the pupil onto the DMs and can also be used to test control algorithms under pupil drift and/or misalignments. A second 350 mm OAP then collimates the beam with the woofer located in the new pupil plane, followed by a pupil relay of two 350 mm OAPs that images the pupil onto the tweeter. The woofer is planned to be an ALPAO DM-97, and the tweeter is an already acquired Boston Micromachines kilo-DM with 34 actuators across its clear aperture.

Both the woofer and the tweeter are placed at a 10$^\circ$ angle with respect to the optical axis, which introduces a small projection effect of the pupil along the $x$-direction: the 12.6 mm pupil produces a 12.8 mm footprint, still comfortably inside both clear apertures. After the tweeter, a set of 870 and 621 mm parent focal length OAPs resizes the pupil to 9 mm, matching the pupil size of MagAO-X in the coronagraphic part of the instrument\cite{males2018}. A beam-splitter is placed after the last XAO OAP to feed the wavefront sensing arm, minimizing non-common-path aberrations. This wavefront sensing arm will be designed at a later stage, with sufficient space reserved to accommodate multiple wavefront sensors.

\subsubsection{Characterization of the Boston Micromachines DM}
\begin{figure}
    \centering
    \includegraphics[width=\linewidth]{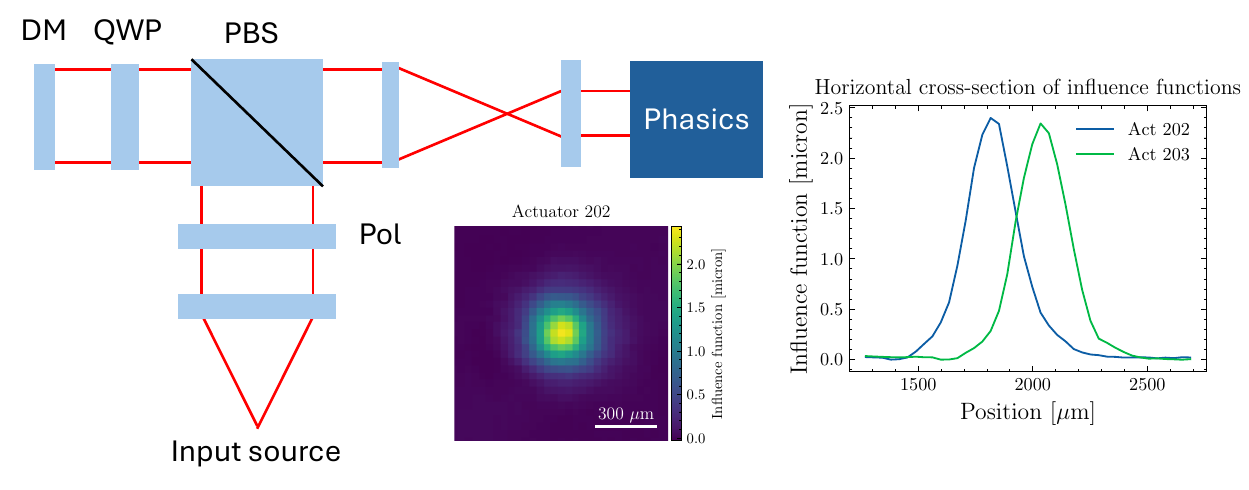}
    \caption{Left: schematic overview of the setup used to characterize and flatten the deformable mirrors, based on a PHASICS lateral shearing interferometer. Middle: measured influence function of a single actuator (actuator 202). Right: horizontal cross-sections through the influence functions of two neighboring actuators (202 and 203), showing their overlap.}
    \label{fig:dm_char}
\end{figure}

We followed a characterization procedure similar to that of van Gorkom et al.\cite{vangorkom2018} to characterize the DMs and obtain a master flat. The setup shown in Fig. \ref{fig:dm_char} is used to characterize the DM and flatten it. It employs a PHASICS lateral shearing interferometer to sense the wavefront with high spatial resolution and high dynamic range. Of the acquired DMs, only the 492-actuator BMC DM has been characterized so far; it is not part of the baseline optical design, but serves as a testbed for the procedure that will be applied to the kilo-DM and the coronagraphic DMs.

A reference phase map was first taken with a fold-mirror at the location of the DM to calibrate out aberrations from the optics and alignment. For the kilo-DM this reference is planned to be taken with a superflat. We then optimized the two-way stroke of the DM following the same procedure, and found a normalized bias voltage of 0.6 V/V$_{max}$ to give the largest two-way stroke.

\begin{figure}
    \centering
    \includegraphics[width=\linewidth]{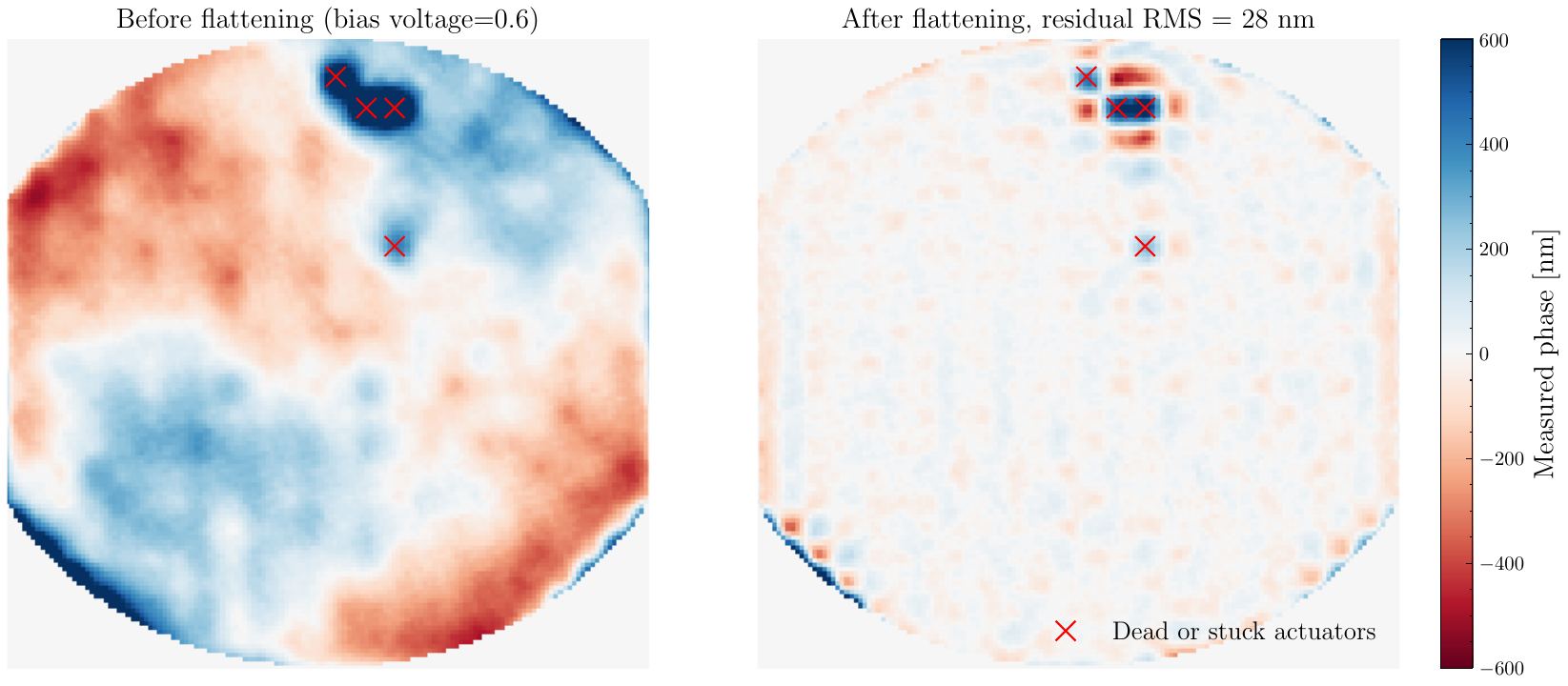}
    \caption{Measured wavefront of the 492-actuator BMC DM before flattening (left) and after flattening (right), relative to the reference fold mirror. The residual after flattening is 28 nm RMS. The red crosses mark dead or stuck actuators.}
    \label{fig:dm_result}
\end{figure}

Fig. \ref{fig:dm_char} shows the measured influence function for one of the actuators and the overlap between two neighboring actuators. Fig. \ref{fig:dm_result} shows the wavefront before and after flattening for the 492-actuator BMC DM, as measured with respect to the fold-mirror. The residual wavefront error clearly shows a number of dead or stuck actuators, and has a 28 nm residual RMS over the aperture after masking the dead actuators. This process will be repeated for all the other acquired DMs.

\subsection{Coronagraph module}
The coronagraphic module provides two spaces for coronagraphic DMs (CDMs), placed right before the apodizer pupil plane. The first CDM spot will initially be populated by a fold mirror, which can later be replaced by a coronagraphic DM (e.g.\ another kilo-DM). The main CDM is currently planned to be a $200\times200$ Fraunhofer deformable mirror with a 10 mm clear aperture. Both CDMs are placed outside the pupil. Although this does not allow pure phase correction with these DMs, this is not necessarily required for dark-hole digging and was deemed preferable over another pupil relay in terms of accumulated wavefront error and physical space. The distance by which the DMs are placed away from each other and from the pupil is a trade-off between the amplitude-correction efficiency, vignetting, effective use of the actuators and physical space \cite{Mazoyer2017SPIE10400E..14M_dm_optimization, baudoz2018}. We will place the future Fraunhofer DM at a distance of 78 mm from the pupil, mainly due to space constraints, while the fold-mirror (which is planned to be upgraded to a second CDM in the future) sits at a distance of 150 mm upstream from this DM.

We have confirmed with the Fresnel and dark-hole digging simulations in Section \ref{sec:fresnel} that for realistic surface errors we still stay well within the stroke limits of the DMs, even for a two-sided dark zone at $\sim10^{-8}$ contrast. Because the Fraunhofer DM has a small relative clear aperture and is not exactly in the pupil plane, it will slightly clip the beam for the largest field angles ($>150\,\lambda/D$), so the full $200\,\lambda/D$ field of view of Table \ref{tab:requirements} is available in the XAO arm but not through the coronagraphic DMs. The only remedy would be an additional pupil relay, but the extra space and accumulated aberrations this would introduce were not deemed worth the small amount of clipping at the largest field angles.

An apodizer filter wheel is placed at the pupil plane to test, for example, phase-apodized vortex\cite{haffert2025_pavvc, landman2026} and Lyot coronagraphs \cite{Por2020ApJ...888..127P}. PIAA lenses can additionally be placed just outside the pupil plane to test PIAA-CMC\cite{2026A&A...708A.144T}, PIAA-Zernike\cite{taras2026b}, or hybrid coronagraph-wavefront sensing masks based on metasurfaces \cite{patel_spie_2026}. The beam is then focused onto the coronagraphic focal plane by a 621 mm OAP, resulting in an F/69 beam. This again matches the MagAO-X design and provides a large enough point spread function, which relaxes the requirements on the smallest features in the manufactured coronagraphic focal-plane masks that are planned to be tested. Light rejected by the focal-plane mask can be re-imaged for focal-plane low-order wavefront sensing (FLOWFS) \cite{Mars2026}. A second 621 mm OAP then collimates the beam and forms a pupil plane where the Lyot stop is placed, with plenty of space to re-image the light rejected by the Lyot stop (LLOWFS) \cite{Singh2015PASP_llowfs}. Refractive lenses will be used to focus the beam onto the science cameras for the respective wavelength ranges. Plenty of space remains for additional science instruments, such as an integral field spectrograph, and for visitor experiments. The bench is designed to explore a range of focal-plane wavefront sensing and dark-hole control strategies, including pair-wise probing\cite{giveon2011}, (implicit) Electric Field Conjugation\cite{giveon2007, haffert2023}, and the self-coherent camera\cite{Baudoz2006_scc, tonucci_spie_2026}.

\subsection{Mechanical design}
\label{sec:mechanical}
All optics will be mounted on an air-floated Thorlabs T1530 optical table ($1.5 \times 3$~m), maintaining a single beam height throughout. A tolerance analysis identified the alignment of the OAPs, and in particular their clocking, as the dominant contributor to the wavefront error. Each OAP will be held in a 2-inch Thorlabs Polaris K2S3 kinematic mount, which provides three-actuator tip/tilt control with a low-distortion flexure retention. The most sensitive degree of freedom, the OAP clocking, is fixed mechanically by registering the manufactured flat on each OAP against a clocking datum in its mount, so that the optic can only be installed at a single rotation. The residual aberrations from realistic misalignments are low-order and well within the stroke of the deformable mirrors. A light-tight, thermally insulated enclosure over the full bench is planned, both for stray-light suppression and for the thermal stability required at these contrast levels.

\section{PERFORMANCE ANALYSIS}
\label{sec:performance}

\subsection{Throughput budget}
\label{sec:throughput}
The largest throughput losses are the pinhole, the aperture stop, and the cumulative reflection losses from the mirrors. The Gaussian beam is spatially filtered at the pinhole and subsequently truncated by the aperture stop, so that the telescope simulator transmits only about 1\% of the incident energy (Fig. \ref{fig:telescope_sim}). This ensures a flat pupil illumination profile at the cost of throughput. The second dominant loss mechanism is reflection, as the optical design includes $\sim$16 mirrors, each assumed to reflect 98\%. Together with the losses in the telescope simulator, these reduce the total throughput to $\sim$0.72\%. Although this is low, it is sufficient for laboratory operation. Combining this throughput with the light source, we expect a photon flux of about $4.8\times10^{-7}$ W/nm at a wavelength of 1 $\mu$m, equivalent to about a $-4.5$ magnitude star observed with the ELT. This comfortably exceeds the minimum photon-flux requirement in Table \ref{tab:requirements}, approaches the goal, and keeps experiment times short.

\subsection{Fresnel analysis \& contrast limits}
\label{sec:fresnel}
The fundamental contrast limit of the testbed is set by a combination of surface errors on the optics, Fresnel propagation effects, and the finite number of DM actuators. We have simulated the accumulation of phase and amplitude errors resulting from Fresnel propagation of the mirror surface errors using \texttt{hcipy}\cite{por2018}, including realistic OAP spacings and surface errors, similar to what was done by N'Diaye et al.\cite{ndiaye2013} for HiCAT. We used the power spectral density (PSD) requirement envelope supplied by the manufacturer of the OAPs to generate worst-case scenario surface errors for the optics. The beam is propagated through the instrument to obtain a coronagraphic image, assuming a second-order ideal coronagraph, and we subsequently use adjoint Electric Field Conjugation based on algorithmic differentiation\cite{2021JATIS...7a9002W, 2025JATIS..11c9001M} to minimize the electric field between 3 and 10 $\lambda/D$. This was necessary because explicitly building and inverting the Jacobian for the $200\times200$ Fraunhofer DM that we have also simulated is very memory-intensive.

\begin{figure}
    \centering
    \includegraphics[width=\linewidth]{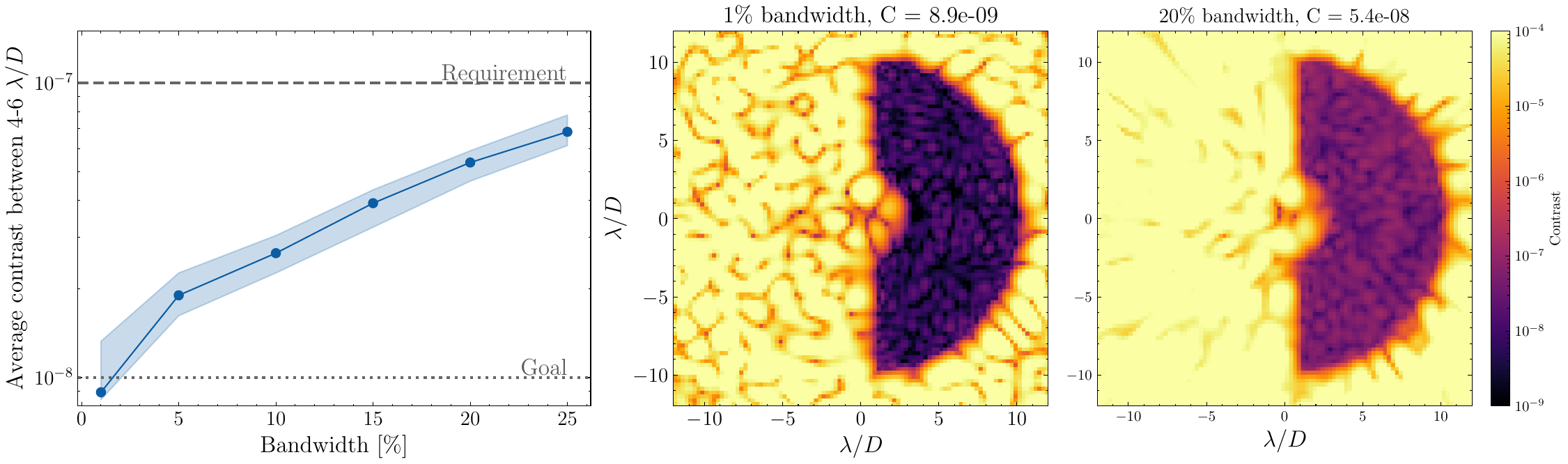}
    \caption{Left: average contrast between 4 and 6 $\lambda/D$ as a function of the spectral bandwidth centered around 800 nm. The shaded band shows the observed spread for random surface-error realizations sampled from the 2D PSD requirement envelope specified by the manufacturer. Right: example focal-plane images after dark-hole digging for a realization of the surface errors in a 1\% and 20\% bandwidth respectively.}
    \label{fig:fresnel}
\end{figure}

Fig. \ref{fig:fresnel} shows the resulting contrast between 4 and 6 $\lambda/D$ as a function of the spectral bandwidth in the ground-based mode centered at 800 nm, together with example focal-plane images after dark-hole digging. For this simulation, we have assumed a single kilo-DM in the pupil plane, the tweeter in this case, for the dark hole digging, as this is the initial situation while a dedicated coronagraphic DM is being procured. We find that the specified surface quality is good enough to reach contrasts of a few times $10^{-8}$ in a 10\% bandwidth at $5\,\lambda/D$, comfortably meeting the $10^{-7}$ requirement and approaching the $10^{-8}$ goal. We note that using manufacturer PSD envelopes provides a conservative baseline, and as-built optics typically exhibit better surface figures.

\begin{figure}
    \centering
    \includegraphics[width=\linewidth]{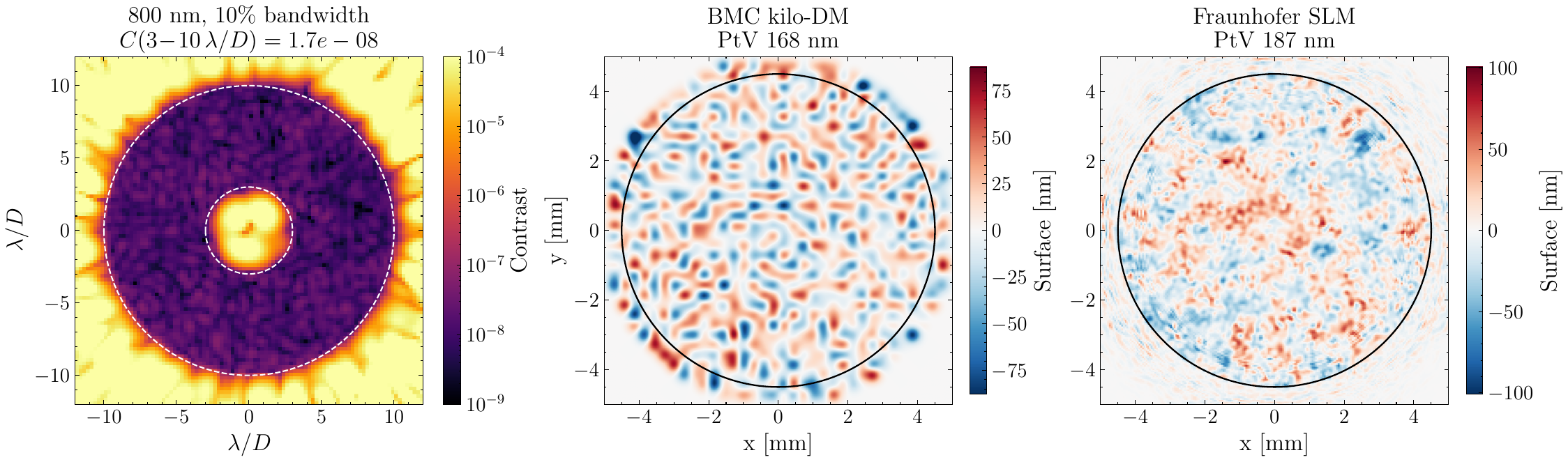}
    \caption{Result of a two-sided dark hole digging simulation with the resulting focal plane image (left) and the final shapes on the two coronagraphic DMs (middle and right).}
    \label{fig:two_sided}
\end{figure}

We have also simulated dark-hole digging with the full planned setup, including two coronagraphic DMs at their physical locations in the testbed. The results for a two-sided dark zone are shown in Fig. \ref{fig:two_sided}, along with the final DM surface, showing that we can reach $\sim 10^{-8}$ two-sided easily within the stroke limit of the DMs. We note that these are of course idealized simulations and that they just consider pessimistic estimates of the surface errors from the OAPs, while in reality other factors may become limiting (e.g. DM quantization, bench stability, EFC model errors). These simulations therefore indicate that the surface quality of the optics are unlikely to be limiting the final contrast.

\section{SOFTWARE \& CONTROL}
\label{sec:software}
A central design driver for NEXT is that algorithms developed and validated on the bench can be transferred to an on-sky instrument with minimal changes. For this reason, the real-time control software will be based on the eXtreme Wavefront Control toolkit\cite{males_spie_2026_xwct}, which was developed for MagAO-X, and relies on the MILK image-processing library and the Compute And Control for Adaptive Optics (CACAO)\cite{guyon2018, deo_spie_2026_cacao} frameworks. These packages have already been deployed on extreme-AO instruments and provide the shared-memory, low-latency infrastructure needed to meet our real-time compute requirements.

The control hardware will be based on commercial off-the-shelf components in a mixed CPU/GPU architecture. This will allow us to also test machine-learning wavefront sensing and control algorithms\cite{2025A&A...696L...1L, nousiainen2022}. Furthermore, the control software is designed for comprehensive telemetry, so that wavefront-sensor measurements, DM commands, and focal-plane images can be recorded synchronously, which is essential for validating control algorithms and forward-modeling algorithms that use multiple cameras\cite{mars_spie_2026}.

\section{STATUS \& OUTLOOK}
\label{sec:outlook}
\begin{figure}
    \centering
    \includegraphics[width=\linewidth]{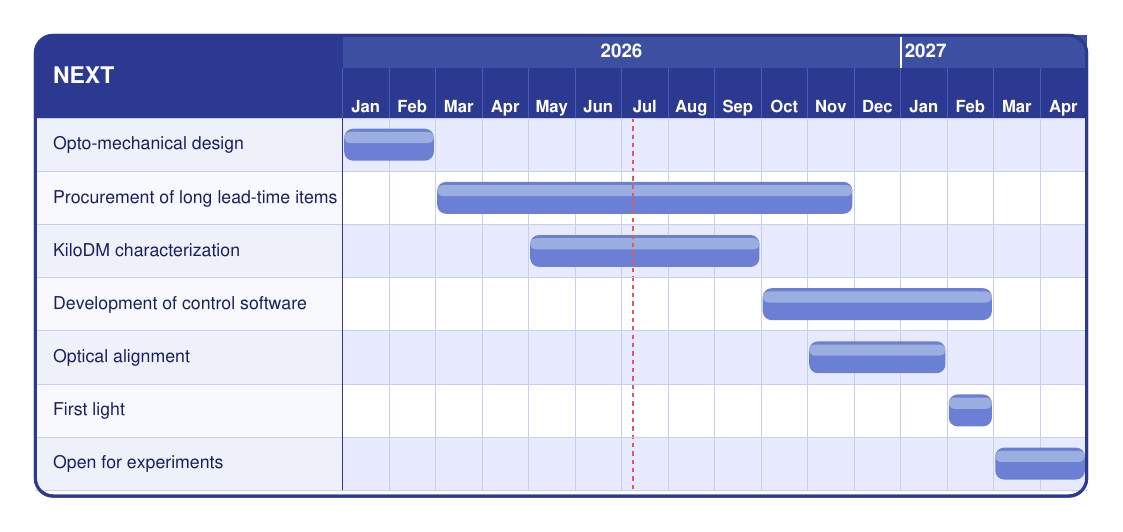}
    \caption{Project schedule for the NEXT testbed, from procurement of long lead-time items through alignment, integration, and the first high-contrast experiments.}
    \label{fig:schedule}
\end{figure}

The opto-mechanical design has been finalized and the long lead-time items have been ordered. We are currently characterizing the deformable mirrors that will go into the setup, and we are awaiting the arrival of the custom OAPs in November 2026. In the meantime, we are setting up the real-time control software and preparing for the alignment and integration of the bench. The overall project schedule is shown in Fig. \ref{fig:schedule}. Once the OAPs are delivered, alignment and integration of the bench are planned for the end of 2026 / start of 2027, using a HeNe laser and the PHASICS sensor. The first high-contrast experiments, closed-loop XAO correction and coronagraphic dark-hole digging, are expected around March/April 2027, targeting the raw-contrast requirement of $10^{-7}$ at $5\,\lambda/D$.

Beyond first experiments, the testbed is designed to grow in capability. The wavefront-sensing arm, for which space has been reserved after the XAO relay, will be designed and populated with one or more Fourier-filtering sensors such as a pyramid wavefront sensor. The fold mirror initially occupying the second coronagraphic DM position is planned to be replaced by a deformable mirror, so that two CDMs enable simultaneous phase and amplitude control. The modular backend leaves room for additional science instruments, including both a visible and a near-infrared camera, as well as for visitor experiments, ranging from new coronagraph and wavefront sensing architectures to novel scientific backends.

\acknowledgments % equivalent to \section*{ACKNOWLEDGMENTS}
The authors acknowledge support from NWO Award 184.036.004.

% References
\bibliography{report} 

\begin{thebibliography}{10}

\bibitem{kasper2021}
{Kasper}, M., {Cerpa Urra}, N., {Pathak}, P., et~al., ``{PCS -- A Roadmap for Exoearth Imaging with the ELT},'' {\em The Messenger}~{\bf 182},  38--43 (Mar. 2021).

\bibitem{males2024gmagaox}
{Males}, J.~R., {Close}, L.~M., {Haffert}, S.~Y., et~al., ``{High-contrast imaging at first-light of the GMT: the preliminary design of GMagAO-X},'' in [{\em Ground-based and Airborne Instrumentation for Astronomy X}{\nolinebreak\hspace{0.1em}]},  {\em Society of Photo-Optical Instrumentation Engineers (SPIE) Conference Series} {\bf 13096},  130960Y (2024).

\bibitem{feinberg2024}
{Feinberg}, L., {Ziemer}, J., {Ansdell}, M., {Crooke}, J., {Dressing}, C., {Mennesson}, B., {O'Meara}, J., {Pepper}, J., and {Roberge}, A., ``{The Habitable Worlds Observatory engineering view: status, plans, and opportunities},'' in [{\em Space Telescopes and Instrumentation 2024: Optical, Infrared, and Millimeter Wave}{\nolinebreak\hspace{0.1em}]},  {Coyle}, L.~E., {Matsuura}, S., and {Perrin}, M.~D., eds., {\em Society of Photo-Optical Instrumentation Engineers (SPIE) Conference Series} {\bf 13092},  130921N (Aug. 2024).

\bibitem{jensenclem2021}
{Jensen-Clem}, R., {Dillon}, D., {Gerard}, B., {van Kooten}, M., {Fowler}, J., {Kupke}, R., {Cetre}, S., {Sanchez}, D., {Hinz}, P., {Laguna}, C., {Doelman}, D., and {Snik}, F., ``{The Santa Cruz Extreme AO Lab (SEAL): design and first light},'' in [{\em Techniques and Instrumentation for Detection of Exoplanets X}{\nolinebreak\hspace{0.1em}]},  {Shaklan}, S.~B. and {Ruane}, G.~J., eds., {\em Society of Photo-Optical Instrumentation Engineers (SPIE) Conference Series} {\bf 11823},  118231D (2021).

\bibitem{jensenclem2025}
{Jensen-Clem}, R., {Chambouleyron}, V., {Javier}, P., {Dillon}, D., {Por}, E.~H., {Calvin}, B., {Cetre}, S., {Amezcua Correa}, R., {Crowe}, T., {Diaz}, J., {Dobias}, C., {Doelman}, D., {Eikenberry}, S., {Fowler}, J., {Gerard}, B.~L., {Hinz}, P., {Kupke}, R., {Moreno}, A., {Nguyen}, T., {Salama}, M., {Sengupta}, A.~R., {Skaf}, N., and {Snik}, F., ``{The Santa Cruz Extreme AO Lab (SEAL) 2.0: A reflective, multi-wavelength rebuild},'' {\em arXiv e-prints} ,  arXiv:2509.03770 (Sept. 2025).

\bibitem{males2018}
{Males}, J.~R., {Close}, L.~M., {Miller}, K., et~al., ``{MagAO-X: project status and first laboratory results},'' in [{\em Adaptive Optics Systems VI}{\nolinebreak\hspace{0.1em}]},  {Close}, L.~M., {Schreiber}, L., and {Schmidt}, D., eds., {\em Society of Photo-Optical Instrumentation Engineers (SPIE) Conference Series} {\bf 10703},  1070309 (2018).

\bibitem{males2022}
{Males}, J.~R., {Close}, L.~M., {Haffert}, S., et~al., ``{MagAO-X: current status and plans for Phase II},'' in [{\em Adaptive Optics Systems VIII}{\nolinebreak\hspace{0.1em}]},  {Schreiber}, L., {Schmidt}, D., and {Vernet}, E., eds., {\em Society of Photo-Optical Instrumentation Engineers (SPIE) Conference Series} {\bf 12185},  1218509 (2022).

\bibitem{males2024}
{Males}, J.~R., {Close}, L.~M., {Haffert}, S.~Y., et~al., ``{MagAO-X: commissioning results and status of ongoing upgrades},'' in [{\em Adaptive Optics Systems IX}{\nolinebreak\hspace{0.1em}]},  {Jackson}, K.~J., {Schmidt}, D., and {Vernet}, E., eds., {\em Society of Photo-Optical Instrumentation Engineers (SPIE) Conference Series} {\bf 13097},  1309709 (2024).

\bibitem{Jovanovic2015PASP..127..890J}
{Jovanovic}, N., {Martinache}, F., {Guyon}, O., {Clergeon}, C., {Singh}, G., {Kudo}, T., {Garrel}, V., {Newman}, K., {Doughty}, D., {Lozi}, J., {Males}, J., {Minowa}, Y., {Hayano}, Y., {Takato}, N., {Morino}, J., {Kuhn}, J., {Serabyn}, E., {Norris}, B., {Tuthill}, P., {Schworer}, G., {Stewart}, P., {Close}, L., {Huby}, E., {Perrin}, G., {Lacour}, S., {Gauchet}, L., {Vievard}, S., {Murakami}, N., {Oshiyama}, F., {Baba}, N., {Matsuo}, T., {Nishikawa}, J., {Tamura}, M., {Lai}, O., {Marchis}, F., {Duchene}, G., {Kotani}, T., and {Woillez}, J., ``{The Subaru Coronagraphic Extreme Adaptive Optics System: Enabling High-Contrast Imaging on Solar-System Scales},'' {\em \pasp}~{\bf 127},  890 (Sept. 2015).

\bibitem{engler2024}
{Engler}, B., {Kasper}, M., {Leveratto}, S., et~al., ``{The GPU-based High-order adaptive OpticS Testbench},'' {\em arXiv e-prints} ,  arXiv:2411.05408 (Nov. 2024).

\bibitem{meeker2021}
{Meeker}, S.~R., {Noyes}, M., {Tang}, H., {Ruane}, G., {Mejia Prada}, C., {Bendek}, E., {Baxter}, W., {Crill}, B., {Riggs}, A.~J.~E., {Poon}, P.~K., and {Siegler}, N., ``{The twin decadal survey testbeds in the high contrast imaging testbed facility at NASA's Jet Propulsion Laboratory},'' in [{\em Techniques and Instrumentation for Detection of Exoplanets X}{\nolinebreak\hspace{0.1em}]},  {Shaklan}, S.~B. and {Ruane}, G.~J., eds., {\em Society of Photo-Optical Instrumentation Engineers (SPIE) Conference Series} {\bf 11823},  118230Y (2021).

\bibitem{noyes2023}
{Noyes}, M., {Walter}, A.~B., {Allan}, G., {Ruane}, G., {Bendek}, E., {Poon}, P.~K., {Mejia Prada}, C., {Riggs}, A.~J.~E., {Tang}, H., and {Meeker}, S., ``{The Decadal Survey Testbed 2: a technology development facility for future exo-Earth observatories},'' in [{\em Techniques and Instrumentation for Detection of Exoplanets XI}{\nolinebreak\hspace{0.1em}]},  {Ruane}, G.~J., ed., {\em Society of Photo-Optical Instrumentation Engineers (SPIE) Conference Series} {\bf 12680},  1268017 (2023).

\bibitem{ndiaye2013}
{N'Diaye}, M., {Choquet}, E., {Pueyo}, L., {Elliot}, E., {Perrin}, M.~D., {Wallace}, J.~K., {Groff}, T., {Carlotti}, A., {Mawet}, D., {Sheckells}, M., {Shaklan}, S., {Macintosh}, B., {Kasdin}, N.~J., and {Soummer}, R., ``{High-contrast imager for complex aperture telescopes (HiCAT): 1. testbed design},'' in [{\em Techniques and Instrumentation for Detection of Exoplanets VI}{\nolinebreak\hspace{0.1em}]},  {Shaklan}, S., ed., {\em Society of Photo-Optical Instrumentation Engineers (SPIE) Conference Series} {\bf 8864},  88641K (2013).

\bibitem{ndiaye2015}
{N'Diaye}, M., {Mazoyer}, J., {Choquet}, {\'E}., {Pueyo}, L., {Perrin}, M.~D., {Egron}, S., {Leboulleux}, L., {Levecq}, O., {Carlotti}, A., {Long}, C.~A., {Lajoie}, R., and {Soummer}, R., ``{High-contrast imager for complex aperture telescopes (HiCAT): 3. first lab results with wavefront control},'' in [{\em Techniques and Instrumentation for Detection of Exoplanets VII}{\nolinebreak\hspace{0.1em}]},  {Shaklan}, S., ed., {\em Society of Photo-Optical Instrumentation Engineers (SPIE) Conference Series} {\bf 9605},  96050I (2015).

\bibitem{ashcraft2022}
{Ashcraft}, J.~N., {Choi}, H., {Douglas}, E.~S., et~al., ``{The Space Coronagraph Optical Bench (SCoOB): 1. design and assembly of a vacuum-compatible coronagraph testbed for spaceborne high-contrast imaging},'' in [{\em Space Telescopes and Instrumentation 2022: Optical, Infrared, and Millimeter Wave}{\nolinebreak\hspace{0.1em}]},  {Coyle}, L.~E., {Matsuura}, S., and {Perrin}, M.~D., eds., {\em Society of Photo-Optical Instrumentation Engineers (SPIE) Conference Series} {\bf 12180},  121805L (2022).

\bibitem{baudoz2018}
{Baudoz}, P., {Galicher}, R., {Potier}, A., {Dupuis}, O., {Thijs}, S., and {Patru}, F., ``{Optimization and performance of multi-deformable mirror correction on the THD2 bench},'' in [{\em Advances in Optical and Mechanical Technologies for Telescopes and Instrumentation III}{\nolinebreak\hspace{0.1em}]},  {Navarro}, R. and {Geyl}, R., eds., {\em Society of Photo-Optical Instrumentation Engineers (SPIE) Conference Series} {\bf 10706},  107062O (2018).

\bibitem{laginja_spie_2026}
{Laginja}, I. et~al., ``{First high-contrast results on THD2 testbed after infrastructure upgrade},'' in [{\em Space Telescopes and Instrumentation 2026: Optical, Infrared, and Millimeter Wave}{\nolinebreak\hspace{0.1em}]},  {\em Proc. SPIE},  to appear (2026).

\bibitem{martinez2022}
{Martinez}, P., {Beaulieu}, M., {Gouvret}, C., {Spang}, A., {Marcotto}, A., {Dejonghe}, J., and {Preis}, O., ``{The segmented pupil experiment for exoplanet detection: 6. from early design to first lights},'' in [{\em Ground-based and Airborne Instrumentation for Astronomy IX}{\nolinebreak\hspace{0.1em}]},  {Evans}, C.~J., {Bryant}, J.~J., and {Motohara}, K., eds., {\em Society of Photo-Optical Instrumentation Engineers (SPIE) Conference Series} {\bf 12184},  121843W (2022).

\bibitem{martinez2023}
{Martinez}, P., {Beaulieu}, M., {Gouvret}, C., {Spang}, A., and {Marcotto}, A., ``{SPEED {\textemdash} Get Ready for the (PCS) Rush Hour},'' {\em The Messenger}~{\bf 190},  55--57 (Mar. 2023).

\bibitem{por2018}
{Por}, E.~H., {Haffert}, S.~Y., {Radhakrishnan}, V.~M., {Doelman}, D.~S., {Van Kooten}, M., and {Bos}, S.~P., ``{High Contrast Imaging for Python (HCIPy): an open-source adaptive optics and coronagraph simulator},'' in [{\em Adaptive Optics Systems VI}{\nolinebreak\hspace{0.1em}]},  {Close}, L.~M., {Schreiber}, L., and {Schmidt}, D., eds., {\em Society of Photo-Optical Instrumentation Engineers (SPIE) Conference Series} {\bf 10703},  1070342 (2018).

\bibitem{vangorkom2018}
{Van Gorkom}, K., {Males}, J.~R., {Close}, L.~M., {Lumbres}, J., {Hedglen}, A.~D., {Schatz}, L., {Kautz}, M., {Guyon}, O., {Knight}, J.~M., {Long}, J.~D., and {Rodack}, A., ``{Characterization of deformable mirrors for the MagAO-X project},'' in [{\em Adaptive Optics Systems VI}{\nolinebreak\hspace{0.1em}]},  {Close}, L.~M., {Schreiber}, L., and {Schmidt}, D., eds., {\em Society of Photo-Optical Instrumentation Engineers (SPIE) Conference Series} {\bf 10703},  1070363 (2018).

\bibitem{Mazoyer2017SPIE10400E..14M_dm_optimization}
{Mazoyer}, J. and {Pueyo}, L., ``{Fundamental limits to high-contrast wavefront control},'' in [{\em Society of Photo-Optical Instrumentation Engineers (SPIE) Conference Series}{\nolinebreak\hspace{0.1em}]},  {Shaklan}, S., ed., {\em Society of Photo-Optical Instrumentation Engineers (SPIE) Conference Series} {\bf 10400},  1040014 (Sept. 2017).

\bibitem{haffert2025_pavvc}
Haffert, S.~Y., Doelman, D.~S., and Landman, R., ``{Phase apodized vortex coronagraphs for arbitrary apertures: a case study for the Habitable Worlds Observatory},'' in [{\em Techniques and Instrumentation for Detection of Exoplanets XII}{\nolinebreak\hspace{0.1em}]},  Ruane, G.~J. and Millar-Blanchaer, M.~A., eds.,  {\bf 13627},  136271B, International Society for Optics and Photonics, SPIE (2025).

\bibitem{landman2026}
{Landman}, R., {Doelman}, D., {Rietjens}, J., {Laginja}, I., {Baudoz}, P., {Peeters}, K., {van Dijk}, C., {Nishie}, Y., {Watanabe}, Y., {Eigenraam}, A., {Vretenar}, M., {van den Born}, J., {Galicher}, R., {Mazoyer}, J., {Potier}, A., {Krasteva}, M., {Taccola}, M., {Bettonvil}, F., and {Snik}, F., ``{SUPPPPRESS: Prototyping and testing liquid-crystal vector vortex coronagraphs with reduced polarization leakage},'' {\em arXiv e-prints} ,  arXiv:2606.10760 (June 2026).

\bibitem{Por2020ApJ...888..127P}
{Por}, E.~H., ``{Phase-apodized-pupil Lyot Coronagraphs for Arbitrary Telescope Pupils},'' {\em \apj}~{\bf 888},  127 (Jan. 2020).

\bibitem{2026A&A...708A.144T}
{Tonucci}, E., {Haffert}, S.~Y., {Foster}, W.~B., {Males}, J.~R., {Guyon}, O., {Close}, L.~M., {Van Gorkom}, K., {Hedglen}, A.~D., {Johnson}, P.~T., {Kautz}, M.~Y., {Kueny}, J.~K., {Li}, J., {Liberman}, J., {Long}, J.~D., {Lumbres}, J., {Mars}, M., {McEwen}, E.~A., {McLeod}, A., {Pearce}, L.~A., {Schatz}, L., and {Twitchell}, K., ``{Phase-Induced Amplitude Apodization Complex Mask Coronagraph (PIAACMC) on-sky demonstration with MagAO-X},'' {\em \aap}~{\bf 708},  A144 (Apr. 2026).

\bibitem{taras2026b}
{Taras}, A.~K., {Haffert}, S.~Y., and {Desdoigts}, L., ``{Differentiable design of the PIAA-ZWFS: a flexible wavefront sensor that approaches the fundamental limit},'' {\em arXiv e-prints} ,  arXiv:2606.28136 (June 2026).

\bibitem{patel_spie_2026}
{Patel}, D. et~al., ``{Demonstration of simultaneous PIAA-coronagraphy and wavefront sensing using a single metasurface-based focal-plane optic},'' in [{\em Advances in Optical and Mechanical Technologies for Telescopes and Instrumentation VII}{\nolinebreak\hspace{0.1em}]},  {\em Proc. SPIE},  to appear (2026).

\bibitem{Mars2026}
{Mars}, M., {Haffert}, S.~Y., {Males}, J.~R., {Close}, L.~M., {Van Gorkom}, K., {Guyon}, O., {Hedglen}, A.~D., {Johnson}, P.~T., {Kautz}, M.~Y., {Kueny}, J.~K., {Landman}, R., {Liberman}, J., {Long}, J.~D., {Lucas}, M., {Lumbres}, J., {McEwen}, E.~A., {McLeod}, A., {Nguyen}, T., {Pearce}, L.~A., {Schatz}, L., {Tonucci}, E., {Twitchell}, K., and {Weinberger}, A.~J., ``{Model-Based Non-Linear Low-Order Wavefront Sensing and Control with a Fully Physical Digital Twin on MagAO-X},'' {\em \aap}  (2026).
\newblock Submitted to Astronomy \& Astrophysics.

\bibitem{Singh2015PASP_llowfs}
{Singh}, G., {Lozi}, J., {Guyon}, O., {Baudoz}, P., {Jovanovic}, N., {Martinache}, F., {Kudo}, T., {Serabyn}, E., and {Kuhn}, J., ``{On-Sky Demonstration of Low-Order Wavefront Sensing and Control with Focal Plane Phase Mask Coronagraphs},'' {\em \pasp}~{\bf 127},  857 (Sept. 2015).

\bibitem{giveon2011}
{Give'on}, A., {Kern}, B.~D., and {Shaklan}, S., ``{Pair-wise, deformable mirror, image plane-based diversity electric field estimation for high contrast coronagraphy},'' in [{\em Techniques and Instrumentation for Detection of Exoplanets V}{\nolinebreak\hspace{0.1em}]},  {Shaklan}, S., ed., {\em Society of Photo-Optical Instrumentation Engineers (SPIE) Conference Series} {\bf 8151},  815110 (2011).

\bibitem{giveon2007}
{Give'on}, A., {Kern}, B., {Shaklan}, S., {Moody}, D.~C., and {Pueyo}, L., ``{Broadband wavefront correction algorithm for high-contrast imaging systems},'' in [{\em Astronomical Adaptive Optics Systems and Applications III}{\nolinebreak\hspace{0.1em}]},  {Tyson}, R.~K. and {Lloyd-Hart}, M., eds., {\em Society of Photo-Optical Instrumentation Engineers (SPIE) Conference Series} {\bf 6691},  66910A (2007).

\bibitem{haffert2023}
{Haffert}, S.~Y., {Males}, J.~R., {Van Gorkom}, K., {Close}, L.~M., {Long}, J.~D., {Hedglen}, A.~D., {Guyon}, O., {Schatz}, L., {Kautz}, M., {Lumbres}, J., {Rodack}, A., {Knight}, J.~M., {Sun}, H., and {Fogarty}, K., ``{Implicit electric field conjugation: Data-driven focal plane control},'' {\em \aap}~{\bf 673},  A28 (May 2023).

\bibitem{Baudoz2006_scc}
{Baudoz}, P., {Boccaletti}, A., {Baudrand}, J., and {Rouan}, D., ``{The Self-Coherent Camera: a new tool for planet detection},'' in [{\em IAU Colloquium 200: Direct Imaging of Exoplanets: Science \& Techniques}{\nolinebreak\hspace{0.1em}]},  {Aime}, C. and {Vakili}, F., eds.,  553--558 (Jan. 2006).

\bibitem{tonucci_spie_2026}
{Tonucci}, E. et~al., ``{Digging dark holes on-sky with the Self-Coherent Camera: Preliminary results},'' in [{\em Adaptive Optics Systems X}{\nolinebreak\hspace{0.1em}]},  {\em Proc. SPIE},  to appear (2026).

\bibitem{2021JATIS...7a9002W}
{Will}, S.~D., {Groff}, T.~D., and {Fienup}, J.~R., ``{Jacobian-free coronagraphic wavefront control using nonlinear optimization},'' {\em Journal of Astronomical Telescopes, Instruments, and Systems}~{\bf 7},  019002 (Jan. 2021).

\bibitem{2025JATIS..11c9001M}
{Milani}, K., {Will}, S.~D., {Van Gorkom}, K., {Douglas}, E.~S., {Ashcraft}, J.~N., and {Cahoy}, K., ``{Demonstrations of adjoint electric field conjugation for a vortex coronagraph},'' {\em Journal of Astronomical Telescopes, Instruments, and Systems}~{\bf 11},  039001 (July 2025).

\bibitem{males_spie_2026_xwct}
{Males}, J.~R. et~al., ``{The eXtreme Wavefront Control Toolkit: High-Contrast Imaging Instrument Control for Ground and Space-Based Coronagraphs},'' in [{\em Software and Cyberinfrastructure for Astronomy IX}{\nolinebreak\hspace{0.1em}]},  {\em Proc. SPIE},  to appear (2026).

\bibitem{guyon2018}
{Guyon}, O., {Sevin}, A., {Gratadour}, D., {Bernard}, J., {Ltaief}, H., {Sukkari}, D., {Cetre}, S., {Skaf}, N., {Lozi}, J., {Martinache}, F., {Clergeon}, C., {Norris}, B., {Wong}, A., and {Males}, J., ``{The compute and control for adaptive optics (CACAO) real-time control software package},'' in [{\em Adaptive Optics Systems VI}{\nolinebreak\hspace{0.1em}]},  {Close}, L.~M., {Schreiber}, L., and {Schmidt}, D., eds., {\em Society of Photo-Optical Instrumentation Engineers (SPIE) Conference Series} {\bf 10703},  107031E (2018).

\bibitem{deo_spie_2026_cacao}
{Déo}, V. et~al., ``{CACAO++: scheduling user experience as the core of adaptive optics real-time computers},'' in [{\em Adaptive Optics Systems X}{\nolinebreak\hspace{0.1em}]},  {\em Proc. SPIE},  to appear (2026).

\bibitem{2025A&A...696L...1L}
{Landman}, R., {Haffert}, S.~Y., {Long}, J.~D., {Males}, J.~R., {Close}, L.~M., {Foster}, W.~B., {Van Gorkom}, K., {Guyon}, O., {Hedglen}, A.~D., {Johnson}, P.~T., {Kautz}, M.~Y., {Kueny}, J.~K., {Li}, J., {Liberman}, J., {Lumbres}, J., {McEwen}, E.~A., {McLeod}, A., {Schatz}, L., {Tonucci}, E., and {Twitchell}, K., ``{Making the unmodulated pyramid wavefront sensor smart: II. First on-sky demonstration of extreme adaptive optics with deep learning},'' {\em \aap}~{\bf 696},  L1 (Apr. 2025).

\bibitem{nousiainen2022}
{Nousiainen}, J., {Rajani}, C., {Kasper}, M., {Helin}, T., {Haffert}, S.~Y., {V{\'e}rinaud}, C., {Males}, J.~R., {Van Gorkom}, K., {Close}, L.~M., {Long}, J.~D., {Hedglen}, A.~D., {Guyon}, O., {Schatz}, L., {Kautz}, M., {Lumbres}, J., {Rodack}, A., {Knight}, J.~M., and {Miller}, K., ``{Toward on-sky adaptive optics control using reinforcement learning. Model-based policy optimization for adaptive optics},'' {\em \aap}~{\bf 664},  A71 (Aug. 2022).

\bibitem{mars_spie_2026}
{Mars}, M. et~al., ``{Forward modelling coronagraphic images with a fully physical, differentiable digital twin of MagAO-X: first laboratory results},'' in [{\em Adaptive Optics Systems X}{\nolinebreak\hspace{0.1em}]},  {\em Proc. SPIE},  to appear (2026).

\end{thebibliography}
\bibliographystyle{spiebib} % makes bibtex use spiebib.bst

\end{document}